\documentstyle[epsfig, twocolumn]{esa_sp_latex}

\begin{document}
%

\parindent 0pt
\parskip 10pt plus 1pt minus 1pt
\hoffset=-1.5truecm
\topmargin=-1.0cm
\textwidth 17.1truecm \columnsep 1truecm \columnseprule 0pt 

\title{\bf {\LARGE A deep search for radio emission from three X-ray 
pulsars :\\
are radio emission and X-ray pulsations anti-correlated ?}}

\author{{\bf R.P.~Fender$^1$, P.~Roche$^1$, G.G.~Pooley$^2$,
D.~Chakrabarty$^3$, A.K.~Tzioumis$^4$, M.A.~Hendry$^1$, 
R.E.~Spencer$^5$} \vspace{2mm} \\
$^1$ Astronomy Centre, University of Sussex, Falmer, Brighton BN1 9QH, U.K.\\
$^2$ MRAO, Cavendish Laborstory, Cambridge CB3 0HE, U.K. \\
$^3$ Centre for Space Research, MIT, Cambridge MA 02139, U.S.A. \\
$^4$ Australia Telescope National Facility, CSIRO, P.O. Box 76, 
Epping 2121, NSW, Australia\\
$^5$ University of Manchester, Jodrell Bank, Macclesfield SK11 9DK, U.K.\\
}

\maketitle

\begin{abstract}

We present results from a deep search for radio emission from the
X-ray pulsar systems GX 1+4, GS 0834-430 \& 4U 0115+63 which
have variously been suggested to possess radio jets and to be
good candidates for propellor ejection mechanisms. None of these
sources is detected at their optical positions, to  
3$\sigma$ limits of a 
few hundred $\mu$Jy. This places upper limits on their radio luminosities 
and thus on the internal energy and numbers of any relativistic
electrons which are three to four orders of magnitude below those
of radio-bright X-ray binaries such as SS 433 \& Cyg X-3. Spectral
and structural information on the proposed `radio lobes' of GX 1+4
make their association with the source unlikely.

The lack of detected radio emission from {\em any} X-ray pulsar
system is discussed statistically, and it is found 
that X-ray pulsations and radio emission from X-ray 
binaries are strongly anti-correlated. \vspace {5pt} \\


\end{abstract}

\section{INTRODUCTION}

Radio emission is observed from $\sim$ 10\% of X-ray binaries
(e.g. Hjellming \& Han 1995). It is generally assumed to be synchrotron 
in nature, characterised by high ($>10^8$ K) brightness temperatures
and negative spectral indices ($\alpha$, where 
$S_{\nu} \propto \nu^{\alpha}$). 
In at least eight cases radio (and in the case of
SS 433, also optical) observations have revealed evidence for an
origin in plasmons ejected from the systems at high velocities
(see e.g. Fender, Bell Burnell \& Waltman 1997 for a review).
X-ray pulsations, indicating the rotation period of a highly magnetised
accreting neutron star, are detected from $\sim 15$\% of X-ray
binaries, mostly of the high-mass variety (e.g. White, Nagase \&
Parmar 1995). 

To date, no radio emission has been detected from an X-ray pulsar
system. In order to investigate whether or not this was simply
bad luck or something more fundamental, we have chosen three
unusual X-ray pulsar systems which might be expected to undergo
significant mass loss, conceivably in the form of a jet:

\begin{itemize}
\item{GX 1+4 is an unusual X-ray binary system containing a
$\sim 2$ min X-ray pulsar
and an M-type giant (hence the `symbiotic X-ray binary' nomenclature)
(Chakrabarty \& Roche 1997), with an unknown
orbital period. The X-ray pulsar
undergoes periods of both spin-up and spin-down as well as 
dramatic changes in luminosity (e.g. Makishima et al. 1988). In the
case of Roche-lobe overflow, it seems that there should be 
significant mass-loss from the system (Chakrabarty \& Roche 1997).
Manchanda (1993) has suggested that two radio hot spots located
roughly symmetrically about the optical position of GX 1+4 may
be lobes resulting from the interaction of radio jets with the
local ISM.}
\item{GS 0834-430 is a 12 sec X-ray pulsar (Aoki et al. 1992)
with a probable orbital period of 110.5 d (Wilson et al. 1997).
No optical counterpart has been identified, so classification
of the system is difficult, but it is probably a Be/X-ray binary
system.}
\item{4U 0115+63 is a well known hard X-ray transient system containing
a 3.6 sec X-ray pulsar in a 24.3 d orbit around a Be star (Rappaport
et al. 1978), prone to large aperiodic outbursts, during which
material may be ejected from the system (e.g. Negueruela et al.
1997).}
\end{itemize}

\section{OBSERVATIONS}

We have observed GX 1+4 \& GS 0834-430 at the same epoch with the
Australia Telescope Compact Array (ATCA) at 20 \& 13 cm, and 
4U 0115+63 at several different epochs with the Ryle Telescope
(RT) at 2 cm.

\subsection{Australia Telescope compact array}

\begin{figure}[p]
  \begin{center}
    \leavevmode
\epsfig{file=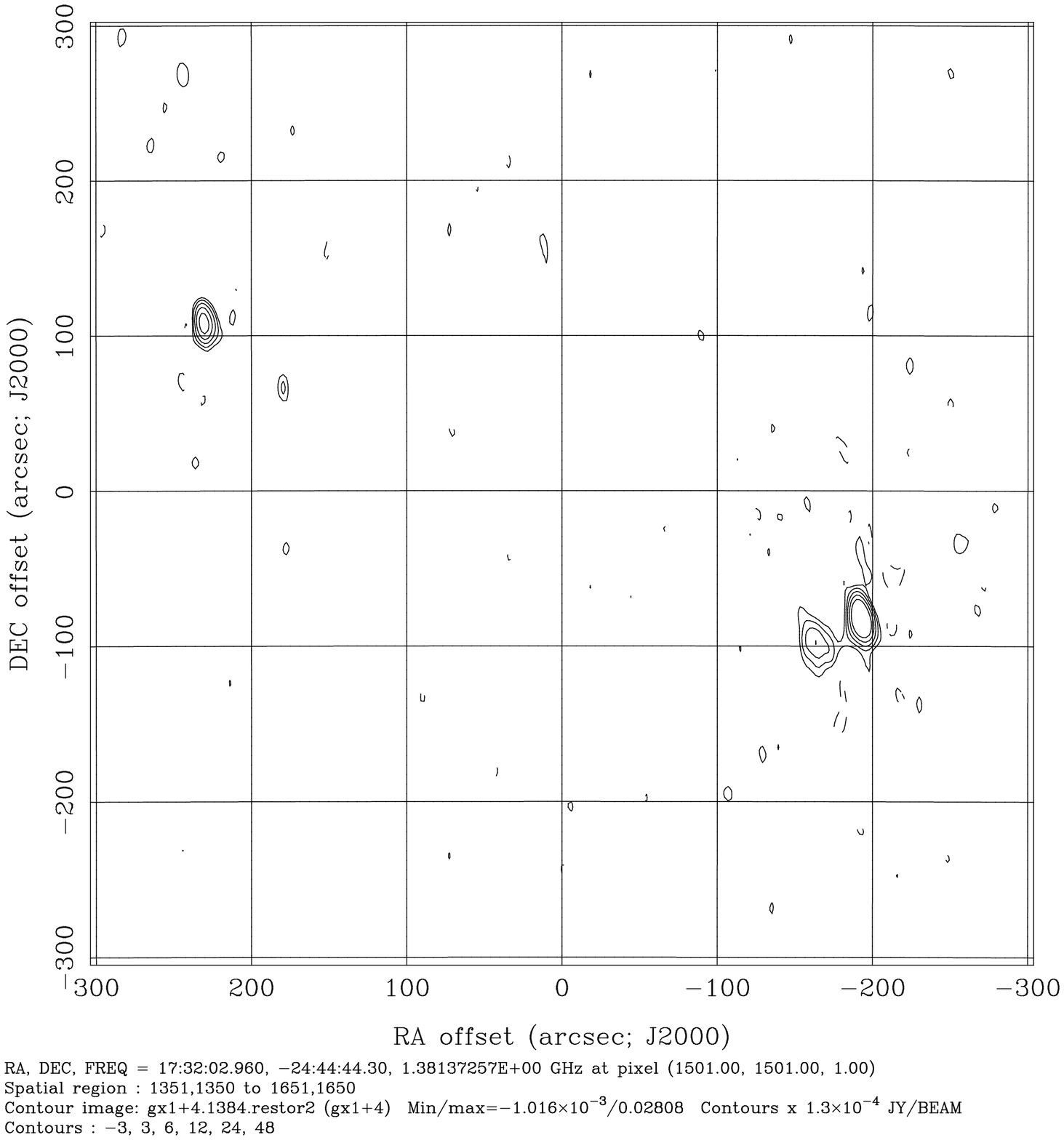, width=8.5cm, bbllx=0pt, bblly=175pt,
	bburx=580pt, bbury=720pt, clip=}
  \end{center}
  \caption{\em 20 cm ATCA map of GX 1+4 field. Contours are at
-3, 3, 6, 12, 24 \& 48 times r.m.s. noise of 130$\mu$Jy. The
sources to the south-west and north-east of the pointing centre
are hot spots `A' \& `B' respectivelyyof Manchanda (1993) --
note that we have resolved `A' into two components.}
\end{figure}

\begin{figure}[p]
  \begin{center}
    \leavevmode
\epsfig{file=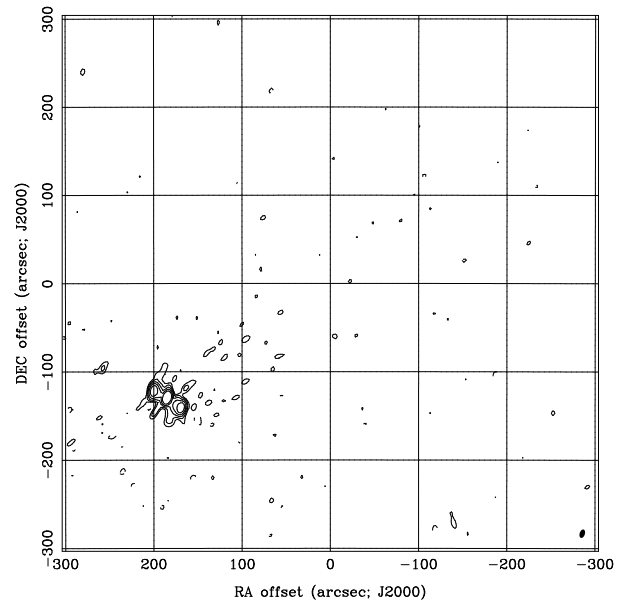, width=8.5cm, bbllx=0pt, bblly=145pt,
	bburx=580pt, bbury=720pt, clip=}
  \end{center}
  \caption{\em 20 cm ATCA map of GS 0834-430 field. Contours are at
-3, 3, 6, 12, 24 \& 48 times r.m.s. noise of 90$\mu$Jy.}
\end{figure}

GX 1+4 \& GS 0834-430 were observed on 1996 May 5 with the ATCA
simultaneously at 20 \& 13 cm. The array was in a compact
configuration but also included the 6 km antenna. Total time on sources
was 8.80 hr for GX 1+4 and 8.45 hr for GS 0834-430. Primary
flux calibration was achieved using 1934-638, and secondary phase
calibrators 1748-253 \& 0823-500 were used for GX 1+4 \& GS 0834-430
respectively. No signal more than 3$\sigma$ above the noise was
detected at the optical position of either source. Maps of the source
fields are presented in Figs 1 and 2 (we refer the reader also to
Mart\'{\i} et al., these proceedings, who reports a marginal detection
of a radio point source at the location of GX 1+4 with the VLA).

\subsection{Ryle Telescope}

Observations with the RT at 2 cm were made at two epochs in 1996.
In the first, we attempted to look for orbitally phase-dependent
transient emission, in particular around periastron. More recently
we made a snapshot observation following BATSE detection of a
brightening of the source. In all observations the primary flux
calibrators were 3C48 and 3C286, and secondary phase calibration
was achieved using B0106+678. Baselines ranged between 72m and
1.2 km.

\begin{figure}[h]
  \begin{center}
    \leavevmode
\epsfig{file=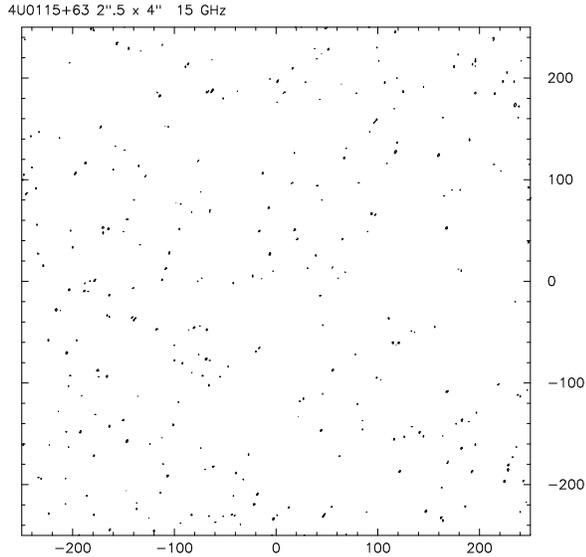, width=8cm, bbllx=0pt, bblly=0pt,
  bburx=600pt, bbury=600pt, angle=270, clip=}
  \end{center}
  \caption{\em 2 cm RT map of 4U 0115+63 field. Contours are at 
-3 \& 3 times the r.m.s. noise of 135 $\mu$Jy. Axes are arcsec 
offsets.}
\end{figure}

\begin{itemize}
\item{{\bf Orbital coverage :}
4U 0115+63 was observed with the RT at 2 cm on the dates of 
1996 Feb 9, 10, 11, 12, 15, 16, 17, 22, 23, 26 \& 28. This
covers phases 0.5 -- 1.4, with periastron occuring around
Feb 21 (based upon periastron passage on JD 2449598.6, and 
P$_{{\rm orb}} = 24.31$ d -- Ignacio Neguerela, private
communication). Individual observations detected no source
to a limit of $\sim 0.8$ mJy; mapping the data gives a
significantly lower limit for the mean flux density (see Fig 3).}
\item{{\bf Observations during outburst :}
Following the announcement by Scott et al. (1996) that 4U 0115+63
had gone into outburst on $\sim$ 1996 Aug 10, we made a further
attempt to observe this source with the RT at 2 cm on 1996 Aug 16.
Again there was no detection, to a 3$\sigma$ limit of 0.8 mJy.}
\end{itemize}

\section{DISCUSSION}

\begin{table*}
\begin{minipage}{177mm}
\centering
\caption{{\em Observed flux limits and derived radio luminosity limits}}
\begin{tabular}{|rcc|cccc|}
\hline
Source & $\lambda$ & 3$\sigma$ flux density & Spectral & Distance & Radio luminosity & Total electron \\
       & (cm)  & limit (mJy) & index $\alpha$   & (kpc) & limit (erg s$^{-1}$) & energy limit (erg) \\
\hline
GX 1+4 &  20 & 0.39 & 0.0 & 6 & $3 \times 10^{30}$ & $9 \times 10^{37}$ \\
       &  13 & 0.24 & -0.6 && $4 \times 10^{30}$ & $5 \times 10^{38}$ \\
\hline
GS 0834-430 & 20 & 0.27 & 0.0 & 7  & $4 \times 10^{30}$ & $1 \times 10^{38}$\\
            & 13 & 0.24 & -0.6 && $5 \times 10^{30}$ & $6 \times 10^{38}$\\
\hline
4U 0115+63  & 2 & 0.44 & 0.0 &5 & $7 \times 10^{30}$ & $2 \times 10^{38}$ \\
            & & & -0.6 && $2 \times 10^{31}$ & $1 \times 10^{40}$\\
\hline
\end{tabular}
\end{minipage}
\end{table*}

\subsection{Limits on synchrotron radio luminosity}

For each source we can evaluate the limits on point-source emission
given the distance to the source and some assumption about the
spectrum of the emission. Based upon observations of radio-bright
X-ray binaries we evaluate the luminosity limits for :
\begin{enumerate}
\item{Optically thin synchrotron emission ($\alpha = -0.6$), implying
plasmon-dominated emission such as that observed in SS433.}
\item{Flat-spectrum core emission ($\alpha = 0.0$), implying significant
absorption at cm wavelengths, such as that observed for Cyg X-3 in
quiescence.}
\end{enumerate}

Assuming isotropic emission, integrated radio luminosity is given by

\[
L = 4\pi d^2 S_0 \nu_0^{-\alpha} \int_{\nu_0}^{\nu_1} \nu^{\alpha}{\rm d}\nu {\phantom{00}}{\rm erg {\phantom{0}}s}^{-1}
\]

where d is distance to source, $\nu_0$ \& $\nu_1$ are lower and upper limits of
integration respectively, and $S_0$ is limit on flux at frequency
$\nu_0$ derived by extrapolating from observed limits. We choose 
$\nu_0 = 5 \times 10^8$ Hz (60cm) and $\nu_1 = 3 \times 10^{11}$ Hz (1mm).

We can go one step further and place limits on the amount of energy in
any relativistic electrons ejected from the sources. Tucker (1977) gives
the following approximation for total energy :

\[
U_R \sim 5 \times 10^{11} B^{-\frac32} L \nu_*^{-\frac12}
\]

where B is the mean plasmon magnetic field (assumed to be randomly aligned),
L is the radio luminosity integrated between frequencies $\nu_0$ \& $\nu_1$,
and $\nu_* = \nu_0$ for spectral index $\alpha > -0.5$ and 
$\nu_* = \nu_1$ for $\alpha \leq -0.5$. 

Table 1 lists observed flux limits, implied radio luminosity 
limits and total relativistic energy limits 
(assuming $B = 0.1$ G -- see e.g. Spencer 1996)  
for both types of radio spectra, for all three sources.
The electron energy limits correspond to $\sim 5 \times 10^{44}$ 
electrons (for optically thin emission), implying a mass limit on the 
ejecta of $\sim 10^{18}$ g for an e$^-$e$^+$ plasma, and $\sim 10^{21}$ g 
for an e$^-$p$^+$ plasma.

For comparison, integrating the radio luminosities of the
compact core ($\alpha \sim 0$) of Cyg X-3 (in quiescence)
and the optically thin emission of SS 433 ($\alpha \sim -0.6$)
between the same limits yields luminosities of $\sim 3 \times 10^{33}$
and $\sim 2 \times 10^{34}$ erg s$^{-1}$ respectively. Thus the three
X-ray pulsars observed here are {\em at least} 500 times less luminous
in the radio regime than radio-bright X-ray binaries. Spencer (1996)
has also calculated $U_R$ for five radio-emitting X-ray binaries and
found that total electron energies are typically $> 10^{41}$ erg,
three orders of magnitude greater than the limits placed here.
Similarly,  Mirabel \& Rodriguez (1994) estimated the mass of a
plasmon (assumed e$^-$p$^+$ composition) ejected from GRS 1915+105
as $10^{25}$ g -- four orders of magnitude more massive than the
limit imposed by our observations.

\subsection{GX 1+4 `jets'}

As well as a limit on point source emission from GX 1+4, we also
investigated hot spots `A' and `B' reported in Manchanda (1993) as
possible radio lobes associated with the source. Whilst the apparent
symmetry of the sources about the optical position of GX 1+4 is
unusual, there is no hint of any collimated emission from
the source to the lobes (Mart\'{\i} et al., these proceedings,
confirms this result). Furthermore, we have resolved Manchanda's
westerly lobe, `A' into two components, aligned roughly East-West,
which have significantly different spectral indices. 

\subsection{Are radio emission and X-ray pulsations anti-correlated ?}

In {\em X-ray Binaries}, van Paradijs (1995) catalogues 193 X-ray binary
systems; White, Nagase \& Parmar (1995) state that there are 32 known
X-ray pulsars, and Hjellming \& Han (1995) list 22 radio-emitting X-ray
binaries. Since this work, $\sim 10$ new X-ray binaries have been found,
at least one of which is an X-ray pulsar (GRO 1744-28 -- no radio 
counterpart), and at least three of which are radio-emitting (GRS 1915+105, 
GRO J1655-40, GRS 1739-278). Furthermore, one previously known X-ray binary 
(GX 339-4) has been found to be radio-bright (see Sood et al., these
proceedings). This gives $\sim 200$ X-ray
binaries, of which $\sim 33$ are X-ray pulsars and $\sim 26$ are 
radio-emitting. And yet radio synchrotron
emission has {\em never been detected from
an X-ray pulsar} (though it should be noted that sensitivity limits
are not uniform across the sample). We do not consider here the
possible radio detection of GX 1+4 as reported by Mart\'{\i}, these
proceedings, as this is both (a) a marginal detection, and (b) 
consistent with thermal emission from the red giant wind in the
system.

We can test the null hypothesis that radio emission and X-ray pulsations 
have no statistical connection against the alternative hypothesis that
characteristics of radio emission and X-ray pulsations are strongly
(anti-)correlated, defining the probability as 

\[
P = C(P_xP_r)^{n_{xr}}[P_x(1-P_r)]^{n_{xr^{\prime}}} [(1-P_x)P_r]^{n_{x^{\prime}r}} 
\]
\[
[(1-P_x)(1-P_r)]^{n_{x^{\prime}r^{\prime}}}
\]

where

\[
C = \frac{n!}{n_{xr}!n_{x^{\prime}r}!n_{xr^{\prime}}!n_{x^{\prime}r^{\prime}}} 
\]

where n is the number in a population, and a subscript of x denotes
X-ray pulsations, r indicates radio emission, and primes indicate
non-membership of the group. We approximate the probability of
being radio or X-ray emitting by their sample estimates, 
$P_x \sim n_x/n$ and $P_r \sim n_r/n$.

We find that the probability of no X-ray pulsars lying in the radio-emitting
X-ray binaries group, {\em if the two properties had no statistical 
connection}, to be $3 \times 10^{-5}$, i.e. if you had $10^5$ samples of
X-ray binaries, you would only expect the observed distribution to occur
$\sim 3$ times. We conclude that radio emission and X-ray pulsations from 
X-ray binaries are statistically anti-correlated.

\section{CONCLUSIONS}

We have performed a deep search for radio emission from three unusual
and interesting X-ray pulsar systems.
However despite deep dual-wavelength imaging of GX 1+4 and GS 0834-430,
and monitoring of 4U 0115+63 around most of an orbit, including periastron,
{\em and} within six days of a recent reported outburst, we find no
evidence for radio emission from these sources. The `radio lobes' claimed
to exist around GX 1+4 are confirmed, but one `lobe' is resolved
into two components, the spectral indices of all three components
differ significantly, and there is no evidence of connecting radio
structure tracing back to the optical location of the source. 

These observations place strong limits on the radio luminosity and
hence numbers, mass, and stored energy of any relativistic ejecta --
three to four orders of magnitude lower than for radio-bright
X-ray binaries. A simple
statistical test shows that the current sample of X-ray binaries and
the subsets of radio-emitting and X-ray-pulsing systems are sufficiently
large that this lack of a radio-emitting X-ray pulsar system cannot
be chance. In fact, it seems that the observable properties of
detectable radio emission and X-ray pulsations from X-ray binaries
are strongly anti-correlated.

We conclude that there appears to be some reason why X-ray pulsars
are not radio-bright, and perhaps more significantly, that X-ray
pulsars cannot form radio jets. This may be due to a lack of a stable
inner accretion disc in systems containing a high magnetic field
neutron star, or the material funnelling down field lines may somehow
disrupt a nascent jet.

\section*{ACKNOWLEDGMENTS}

We thank Ignacio Negueruela and Josep Mart\'{\i} for useful discussions.

\end{document}